\documentclass[epj]{svjour}
\usepackage{amsmath}
\usepackage{mathtools}
\usepackage{algorithm}
\usepackage{algorithmic}
\usepackage{enumerate}
\usepackage{enumerate}
\usepackage{enumerate}
\usepackage{graphics}
\usepackage{amssymb}
\usepackage{mathtools,cuted}
\usepackage{color}
\usepackage{lipsum}
\usepackage{booktabs}
\usepackage[square, comma, sort&compress, numbers]{natbib}
\usepackage{dcolumn}
\usepackage[figuresright]{rotating}
\usepackage[colorlinks=true,linkcolor=blue, citecolor=blue]{hyperref}

\begin{document}
	\title{Critical behavior of a stochastic anisotropic Bak-Sneppen model}
	
	\author{Jihui Han\thanks{\email{hanjihui@zzuli.edu.cn}}\inst{1} \and Wei Li\thanks{\email{liw@mail.ccnu.edu.cn}}\inst{2} \and Zhu Su\inst{3} \and Webing Deng\inst{2}}

	\institute{School of Computer and Communication Engineering, Zhengzhou University of Light Industry, Zhengzhou, P.R. China \and Complexity Science Center, Central China Normal University, Wuhan, P.R. China \and National Engineering Laboratory for Technology of Big Data Applications in Education, Central China Normal University, Wuhan, P.R. China}
	%
	%
\abstract{
In this paper we present our study on the critical behavior of a stochastic anisotropic Bak-Sneppen (saBS) model, in which a parameter $\alpha$ is introduced to describe the interaction strength among nearest species. We estimate the threshold fitness $f_c$ and the critical exponent $\tau_r$ by numerically integrating a master equation for the distribution of avalanche spatial sizes. Other critical exponents are then evaluated from previously known scaling relations. The numerical results are in good agreement with the counterparts yielded by the Monte Carlo simulations. Our results indicate that all saBS models with nonzero interaction strength exhibit self-organized criticality, and fall into the same universality class, by sharing the universal critical exponents.
} 
\maketitle
\section{Introduction}
\label{intro}
Many natural and social phenomena appear to evolve intermittently with bursts, rather than smoothly and gradually. For instance, earthquakes \cite{GR1956}, economic activity,  and biological evolution \cite{SJG1977, doi:10.1142/S0129183101001857}, these systems are far from equilibrium and therefore fine tuning of specific parameters is rare and unlikely. To understand the origin of this ubiquitous spatiotemporal complexity in nature, Bak et al. \cite{PhysRevLett.59.381} proposed self-organized criticality (SOC), namely, systems that are far from equilibrium evolve through many transient states (which are not critical) to a dynamical attractor poised criticality.

As one of the mechanisms of complexity in nature, SOC has become a well studied concept in non-equilibrium statistical mechanics \cite{Bak18071995,DICKMAN2000, PhysRevE.57.5095}.  In the past few decades, a variety of simple models with extremal dynamics that exhibit SOC have been introduced and analyzed, including the Bak-Sneppen (BS) model \cite{Bak1993}, the Sneppen interface depinning model \cite{PhysRevLett.69.3539}, and the Zaitsev model \cite{Zaitsev1992411}. In these models, the system is driven by sequentially updating the site with globally extremal value and such information is propagated throughout the system via local interactions. These models, representing different universality classes, are nevertheless similar to invasion percolation \cite{0305-4470-16-14-028, PhysRevA.45.R8309, PhysRevE.49.1238, PhysRevLett.70.3832}. Among the SOC models, the BS model which mimics the biological evolution of an ecology of interacting species is by far the simplest one.

In the BS model, an ecosystem is characterized by $L^d$ species on a $d$-dimensional lattice with linear size $L$. A random number $f_i$, drawn from the uniform distribution between 0 and 1, is assigned to each site of the lattice as its initial fitness. At each time step, the site with the smallest fitness and its $2d$ nearest neighbors are replaced by new random numbers drawn from the same distribution. After some transient time (which depends on the system size) the model can reach a statistically stationary state where the fitness is uniformly distributed in the range of $f_c$ to 1 and vanishes below $f_c$, where $f_c$ is the threshold fitness. When the stationary state  is reached, the model exhibits punctuated equilibrium and the complexity of this regime can be revealed by the existence of a power-law distribution of avalanche sizes. A critical avalanche is a sequence of successive mutation events with the global smallest fitness below $f_c$. The lifetime $s$ of an avalanche is the total number of events during the avalanche, which obeys a power-law distribution $P(s)\sim s^{-\tau}$ when the system size becomes infinitely large. The BS model has been studied on hypercubic lattices, and the values of $f_c$ and $\tau$ are found to be 0.67702(8) and 1.073(3) \cite{Grassberger1995277} for $d=1$, and 0.328855(4) and 1.245(10) \cite{PhysRevE.53.414, PhysRevE.62.295} for $d=2$, respectively. The upper critical dimension (where $\tau$ reaches its mean-field value 1.5) has been argued to be 4 \cite{PhysRevLett.84.2267} or 8 \cite{PhysRevLett.80.5746}.

Since this extremely rich dynamic critical behavior arises out of truly minimalistic dynamical rules, the BS model has received much attention in the statistical physics community and has been studied through various approaches, including numerical simulations \cite{Grassberger1995,PhysRevE.53.414, Garcia2004164}, theoretical analysis \cite{Maslov1996, doi:10.1142/S0129183101002711}, and mean-field theory \cite{PhysRevLett.71.4087, Marsili1998}. Many variants of the original BS model have been proposed: (1) the discrete BS model in which fitnesses only take the values $0$ and $1$; (2) the anisotropic BS (aBS) model \cite{PhysRevE.58.7141} in which only the least fit species and its right nearest neighbor are mutated; (3) the stochastic BS model \cite{PhysRevE.80.021132} in which the species with the smallest fitness and only one randomly selected from the nearest neighbors are updated; (4) the generalized BS model \cite{0256-307X-19-10-307,LW} in which a parameter $\alpha$ is introduced to describe the interaction strength among nearest species; (5) the random neighbors BS model \cite{PhysRevE.60.R1111} in which the smallest fitness and $K-1$ randomly chosen other ones are updated. Variants of the BS model with exponentially and power-law distributed random numbers have also been studied \cite{PhysRevLett.75.1969, PhysRevE.58.3993}.

It has been observed that increasing the size of the mutation zone affects the threshold fitness $f_c$ but not critical exponents \cite{Garcia2004516, PhysRevE.80.021132}, and the isotropic and anisotropic mutation zones have different critical exponents \cite{PhysRevE.58.7141}. In Ref.~\cite{PhysRevE.53.414}, the critical exponents of the BS model were related by a scaling theory, and only two of them characterize the BS model. In Ref.~\cite{Marsili1998}, the avalanche hierarchy equation yields a new relation between exponents in the BS model, thus reducing the number of independent exponents to just one. In Ref.~\cite{PhysRevE.84.041124} the BS model was connected to more tractable Markovian processes, and in the case of a large number of species, the long-time behavior of the fitness profile in the BS model can be replicated by a model with a purely rank-based update rule whose asymptotics can be studied rigorously. Moreover, in Refs.~\cite{PhysRevE.61.771, PhysRevE.61.5630, PhysRevE.62.7743} a different hierarchy of avalanches was observed by introducing the average fitness which may be a good quantity in determining the emergence of criticality. The convergence dynamics (both short-time and long-time) of the BS model have been analyzed in Refs.~\cite{doi:10.1142/S0129183103004942, Bakar20085110, Tirnakli2004151}. The BS model has been studied on different heterogeneous graphs as well, e.g., random networks \cite{PhysRevLett.81.2380}, adaptive networks \cite{ReviewOfAdaptiveCoevolutionaryNetworks, 2007/11//print, refId0, Paczuski2004158} , small-world networks \cite{1999cond.mat..5066K} and scale-free networks \cite{0295-5075-57-5-765}. The threshold fitness approaches zero as the scale-free network size increases to infinity.

In this paper we introduce and study a stochastic version of the original anisotropic Bak-Sneppen (saBS) model. The stochasticity is introduced by randomly selecting the sites of the mutation zone. We estimate the threshold fitness and the critical exponents using two different methods: master equation and Monte Carlo simulation. In the next section, we briefly describe the mechanisms of the saBS model. In Sec.~\ref{Results} we present the main results and findings. Conclusions are given in Sec.~\ref{Conclusions}

\section{Model description}
\label{sec:1}
The saBS model is defined as follows:
\begin{enumerate}[(1)]
	\item $L^d$ species are located on a $d$-dimensional lattice of linear size $L$. Initially, $L^d$ random numbers drawn from the distribution $P(f)=e^{-f}$ are assigned independently to each species as fitness.
	\item \label{step2} At each time step, the smallest fitness in the system will be replaced with a new random number from $P(f)$. The fitness of its $d$ right nearest neighbors (located in positive directions of corresponding coordinates) will be replaced with random numbers also drawn from $P(f)$, but with probability $\alpha$.
	\item Repeat (\ref{step2})
\end{enumerate}

Here $\alpha$ can be interpreted as interaction strength. It is a fixed parameter during the evolution. If $\alpha$ is set to 0, there is no interaction and eventually all fitness values approach infinity. In this case, no SOC can be observed. If $\alpha$ is set to 1, the original aBS model is restored.

\section{Results}
\label{Results}
\subsection{The results from master equation}	
Our results of the $1d$ saBS model are based on an exact equation for the probability distribution function $Q(r,f)$ of spatial sizes $r$ (characteristic length of an avalanche) of $f$-avalanches. An $f$-avalanche is defined as a sequence of successive mutation events with the smallest fitness $f_{min} < f$. This master equation was first introduced in Ref.~\cite{PhysRevE.58.7141}. Let us recall briefly the sequence of logical steps leading to this equation. The starting point is the analysis of how $Q(r,f)$ changes when $f$ is increased by an infinitesimal amount $df$. Some avalanches of spatial size $r$ will merge with the next one. This event can only occur if at least one of the $r$ sites has a number $f_i<f$, which is randomly drawn from an exponential distribution $\mathcal{P}(f)=e^{-f}$ during the avalanche. When this avalanche stops, according to the definition of $f$-avalanche, all these $r$ sites have $f_i>f$. We can therefore regard the $f_i$'s on these sites as randomly drawn from an exponential distribution normalized between $f$ and $\infty$. The probability that a particular $f_i$ being in the interval $[f, f+df]$ is just $df$. The probability that at least one of the $r$ sites has an $f_i$ in the interval $[f,f+df]$ is $rdf+O(df^2)$. This implies that the number of $f$-avalanches which will merge with the next one when $f$ reaches to $f+df$ is $dQ(r)|_{loss}=-rdfQ(r,f)+O(df^2)$.

Let us now consider a merging event between two $f$-avalanches of spatial sizes $r_1$ and $r_2$ resulting in a $f+df$ avalanche of size $r$. There are two scenarios of how this can happen.	
\begin{enumerate}
	\item The rightmost point of the second avalanche is at a displacement $r$ from the leftmost point of the first avalanche; the constraint on possible values of $r_1$ and $r_2$ imposed by this scenario is $max(r_1,r_2)\le r\le r_1+r_2-1$.
	\item $r_1=r$, and the second avalanche is fully contained within the first one.
\end{enumerate}
In the former case, the values of $r$, $r_1$, and $r_2$ uniquely specify the initial site of the second avalanche. Therefore, the probability of this event occurring is just the probability $df+O(df^2)$ that this site has $f_i\in [f, f+df]$. However, in the latter case the starting point of the second avalanche can be any of the first $r-r_2$ sites of the first avalanche. This event occurs with a probability $(r-r_2)df+O(df^2)$. Putting all these terms together, we find		
\begin{align}
\partial_f Q(r,f)=&-rQ(r,f)+\sum_{r_1=1}^r Q(r_1,f)\sum_{r_2=r-r_1+1}^r Q(r_2,f) \notag \\
&+Q(r,f)\sum_{r_2=1}^r(r-r_2)Q(r_2,f).
\label{AHE}
\end{align}	
This is an exact equation for the distribution of spatial sizes of $f$-avalanches. Its validity does not require any scaling assumptions.

Clearly, Eq.~(\ref{AHE}) for $Q(r,f)$ involves only $Q(r',f)$ with $r'\le r$. Therefore, in principle this distribution can be computed numerically for $r\le R$ to the desired accuracy. In the aBS model Eq.~(\ref{AHE}) has to be solved with the initial condition
\begin{equation}
Q(r,f=0)=\delta_{r,2}.
\label{init1}
\end{equation}	
Whereas, in the 1-d saBS model, the initial condition shall be changed to,	
\begin{equation}
Q(r,f=0)=(1-\alpha)\delta_{r,1}+\alpha\delta_{r,2}.
\label{init2}
\end{equation}
Because in the 1-d saBS model, when $f=0$, $r=2$ sites are updated with probability $\alpha$, and $r=1$ site is updated with probability $1-\alpha$.

\begin{figure}[hbpt]
	\begin{center}
		\includegraphics[width=\linewidth]{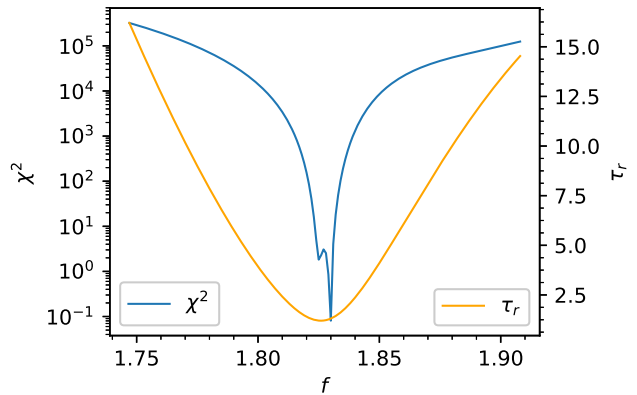}
	\end{center}
	\caption{(Color online) The plots of $\chi^2$ and $\tau_r$ versus $f$ for $\alpha=0.5$ and $r_{max}=4096$. By minimizing $\chi^2$, we get $f_c=3.302$ and $\tau_r=1.298$. $Q(r,f)$ was obtained by numerical integration of Eq.~(\ref{AHE}) with $r\le r_{max}=4096$. A second-order Runge-Kutta method with $\delta f=0.001$ was used.}
	\label{chitaur01}
\end{figure}

\begin{figure}[h]
	\begin{center}
		\includegraphics[width=\linewidth]{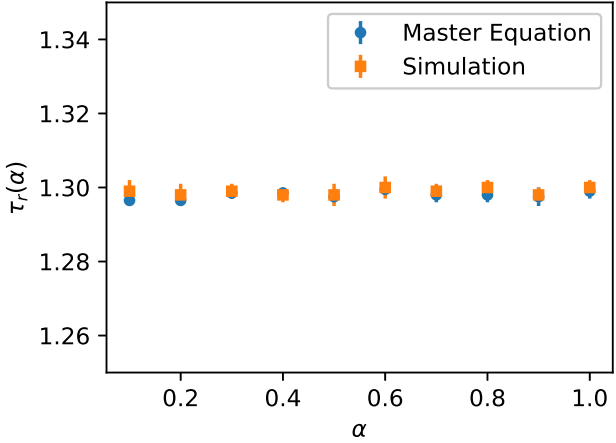}
	\end{center}
	\caption{(Color online) Dependence of exponent $\tau_r$ on $\alpha$.}
	\label{taur}
\end{figure}

The solution of Eq.~(\ref{AHE}) shows that when $f$ is at its critical value $f_c$, $Q(r,f)$ develops a power law relation $Q(r,f)\sim r^{-\tau_r}$. In order to locate the critical point $f_c$, we numerically integrated Eq.~(\ref{AHE}) forward in $f$  with the initial condition Eq.~(\ref{init2}) for several values of $\alpha$, and a least square fit of
$\log Q(r,f)$ versus $\log r$ was performed runtime for each value of $f$. The value $\chi^2(f)$ of the sum of the squared distances from the fit drops nearly to zero in a very narrow region (see Fig.~\ref{chitaur01}), which allows for a very precise estimate of $f_c(\alpha)$ and $\tau_r(\alpha)$. The results of $f_c(\alpha)$ and $\tau_r(\alpha)$ are shown in Fig.~\ref{fcalpha} and \ref{taur} respectively.

\subsection{The results from numerical simulations}
Values of the basic parameters are: $d=1$, $L$ increases from $2^7$ to $2^{14}$ by a factor of 2, $\alpha$ increases from 0 to 1 in step of $0.1$. For each $\alpha$, we ran 100 simulations with different initial conditions and most results obtained are averaged over 100 different realizations. Searching for the minimal fitness using a brute-force algorithm needs to check the fitness values of all species, which requires time $O(L)$. In order to speed up the simulation process, a minimum binary heap is used to store all the fitness. This reduces the time complexity of searching for the minimal fitness from $O(L)$ to $O(\log(L))$.

\subsubsection{Threshold fitness}
When the system reaches the critical state, almost all the fitness will be higher than $f_c$, and the fitness distribution is as follows:
\begin{equation}
\label{Pf}
P(f) = \begin{cases}
e^{-(f-f_c)} & f\ge f_c \\
0  & 0\le f <f_c
\end{cases}
\end{equation}

\begin{figure}[hbpt]
	\begin{center}
		\includegraphics[width=\linewidth]{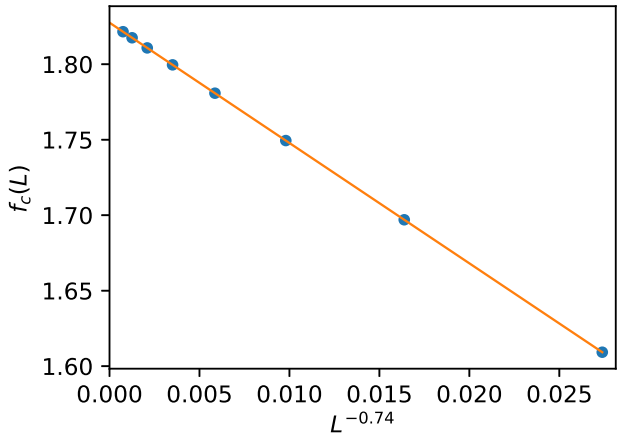}
	\end{center}
	\caption{The threshold fitness values for different system sizes in the case of $\alpha=0.5$. The values of $f_c(L)$ are plotted with $L^{-0.74}$ and extrapolated to obtain $f_c(\infty)=1.830$. The solid line is a least square fit of the data.}
	\label{fcL05}
\end{figure}

\begin{figure}[h]
	\begin{center}
		\includegraphics[width=\linewidth]{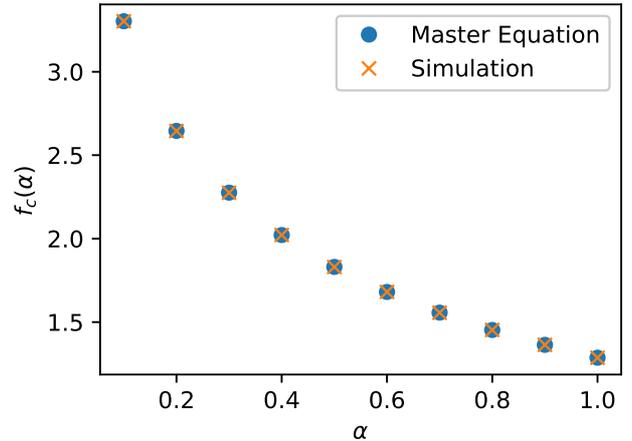}
	\end{center}
	\caption{Dependence of threshold fitness $f_c$ on $\alpha$.}
	\label{fcalpha}
\end{figure}

In $L\rightarrow \infty$ limit the average fitness $\langle f(\infty)\rangle=\int_{0}^{\infty}fP(f)df=1+f_c(\infty)$, which gives $f_c(\infty)=\langle f(\infty)\rangle-1$. The whole fitness profile is sampled at the interval of every $L$ mutation events and $\langle f(\infty)\rangle$ is averaged over $10^6$ fitness profiles. The value of $f_c(\infty)$ is estimated by extrapolating $f_c(L)$ values with $L^{-\kappa}$ (see Fig.~\ref{fcL05}). The most suitable value of $\kappa$ is obtained by minimizing the fitting error. Fig.~\ref{fcalpha} shows the values of $f_c(\infty)$ for different interaction strengths. Clearly, $f_c(\infty)$ decreases as $\alpha$ increases. The threshold values obtained from direct numerical integration of Eq.~\ref{AHE} are in good agreement with those results from the Monte Carlo simulations.

\subsubsection{Avalanche size distribution}
A critical avalanche is defined as a sequence of successive events for which the smallest fitness $f_{min}< f_c$ is confined between two events when $f_{min}>f_c$. The life time $s$ of the avalanche is the total number of successive events in the avalanche, and the spatial size $r$ of the avalanche is the number of species affected by the avalanche. For each system size and interaction strength, we counted 10 million critical avalanches to study their statistical properties.

\begin{figure}[hbpt]
	\begin{center}
		\includegraphics[width=\linewidth]{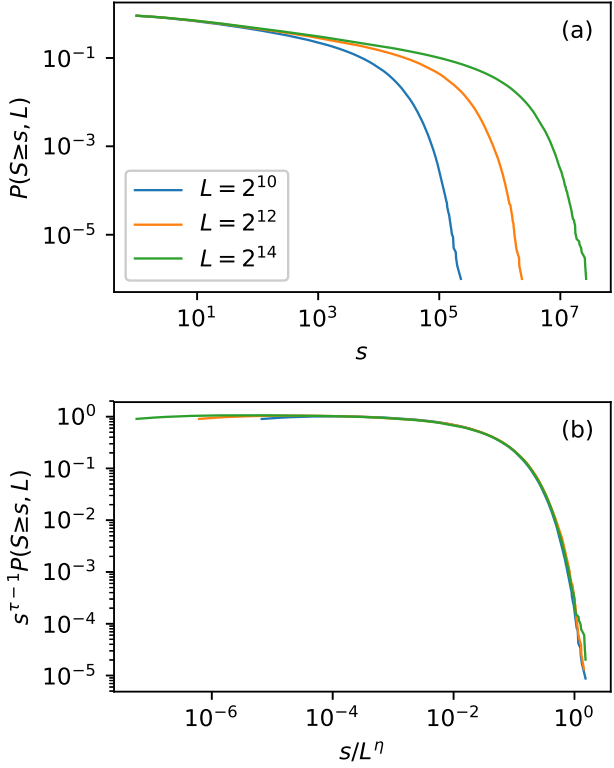}
	\end{center}
	\caption{(Color online) The CCDF $P(S\ge s, L)$ for the $1d$ saBS model with $\alpha=0.5$. In (a) we show the plots for the system sizes $L=2^{10},\ 2^{12},\ 2^{14}$. (b) A finite-size scaling of this data shows an excellent data collapse for $\tau=1.176$ and $\eta=1.719$.}
	\label{PsL05}
\end{figure}

\begin{figure}[hbpt]
	\begin{center}
		\includegraphics[width=\linewidth]{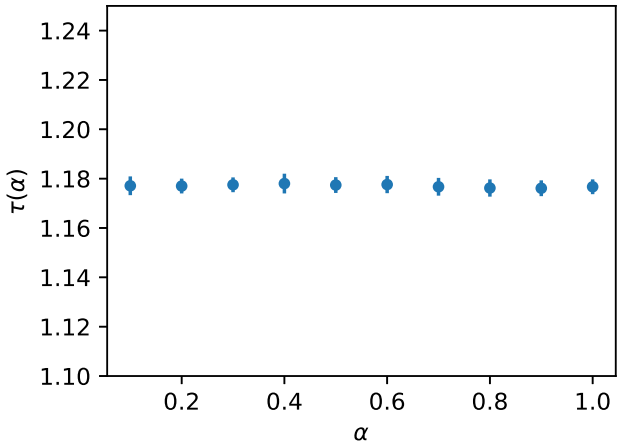}
	\end{center}
	\caption{Dependence of exponent $\tau$ on $\alpha$.}
	\label{taualpha}
\end{figure}

Fig.~\ref{PsL05}(a) shows the complementary cumulative distribution (CCDF) $P(S\ge s, L)$ for $L=2^{10},\ 2^{12}, \ 2^{14}$ with $\alpha=0.5$. Clearly, the statistics of critical avalanche depends on the value of $L$. The larger the system size, the larger the expected size of the avalanche. A finite-size scaling analysis has been done using the following scaling form
\begin{equation}
\label{scalingform}
P(S\ge s, L)\propto s^{-\tau+1}g(s/L^\eta).
\end{equation}
Here, $\tau$ is the life time distribution exponent of critical avalanches. $g(x)$ is a scaling function which decays rapidly to zero for $x\gg 1$ and goes to a constant in the limit of $x\rightarrow 0$. The exponents $\tau$ and $\eta$ fully characterize the scaling of $P(S\ge s, L)$. Therefore, the values of $P(S\ge s, L)$ for various $s$ and $L$ can be collapsed onto a single curve if $s^{\tau -1}P(S\ge s, L)$ is plotted against $s/L^\eta$. The exponents $\tau$ and $\eta$ can be obtained from the best data collapse. In order to remove the subjectiveness of the data collapse, we adopted the method from Ref.~\cite{0305-4470-34-33-302}. Fig.~\ref{PsL05}(b) shows the data collapse for $\alpha=0.5$. Data collapse for other values of $\alpha$ have similar results, and the values of exponent $\tau$ for different $\alpha$ are very close to each other and are roughly equal to 1.176 (see Fig.~\ref{taualpha}).

\begin{figure}[hbpt]
	\begin{center}
		\includegraphics[width=\linewidth]{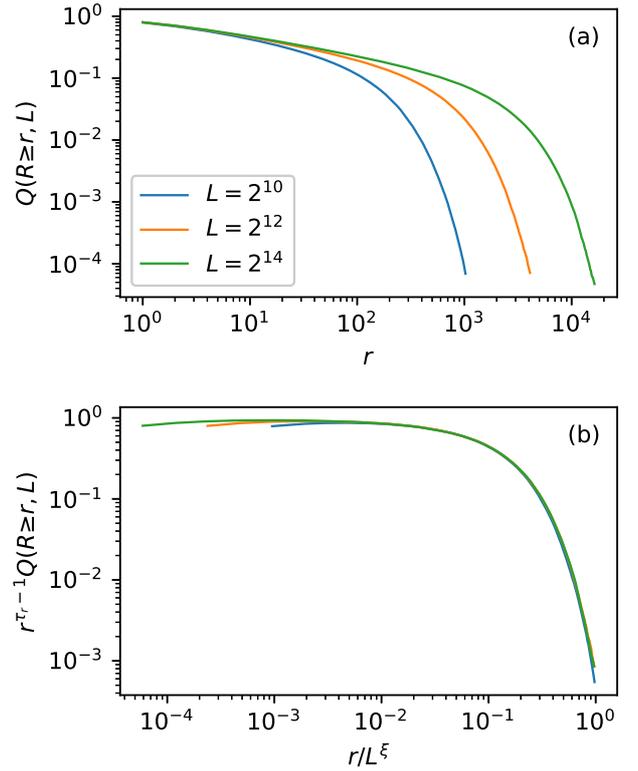}
	\end{center}
	\caption{(Color online) The CCDF $Q(R\ge r, L)$ for the $1d$ saBS model with $\alpha=0.5$. In (a) we show the plots for the system sizes $L=2^{10},\ 2^{12},\ 2^{14}$. (b) A finite-size scaling of this data shows an excellent data collapse for $\tau_r=1.299$ and $\xi=1.003$.}
	\label{QrL05}
\end{figure}

Analogously, for avalanche spatial size, we assume the following scaling ansatz
\begin{equation}
\label{scalingformr}
Q(R\ge r, L)\propto r^{-\tau_r+1}h(s/L^\xi)
\end{equation}
where $\tau_r$ is the critical exponent for the avalanche spatial size distribution , and $h(x)$ is a scaling function similar to $g(x)$. Fig.~\ref{QrL05}(a) shows the CCDF $Q(R\ge r, L)$ for $L=2^{10},\ 2^{12},\ 2^{14}$ in the case of $\alpha=0.5$, and Fig.~\ref{QrL05}(b) shows the data collapse. The values of $\tau_r$ obtained from the data collapse are very close to the counterparts extracted from Eq.~(\ref{AHE}) within numerical uncertainty. $\tau_r$ roughly equals 1.299 for different $\alpha$, which is consistent with the result of the aBS model \cite{PhysRevE.58.7141}.

The above results show that $\tau$ and $\tau_r$ are insensitive to the changes of $\alpha$. This implies the critical behavior of the saBS model is nearly independent of the interaction details. These results are in good agreement with the numerical integration of Eq.~(\ref{AHE}).

\subsection{Scaling relations}	
The moment $\langle r\rangle$ of the distribution $Q(r,f)$ as a function of $f$ can be described by \cite{PhysRevE.58.7141}
\begin{equation}
\partial_f\langle r\rangle = -\langle r^2\rangle +\sum_{\substack{r_1=1\\ r_2=1}}^\infty\frac{ \left[r_1^2+r_2^2+r_1r_2-\min \left(r_1,r_2\right)\right]}{2Q\left(r_1,f\right)Q\left(r_2,f\right)}.
\end{equation}	
For $f<f_c$, when there are no infinite avalanches and the avalanche spatial size distribution $Q(r,f)$ is normalized to 1, one gets	
\begin{equation}
\partial_f\langle r\rangle =\frac{\langle r\rangle^2}{2}-\frac{\langle \min \left(r_1, r_2\right)\rangle}{2}.
\end{equation}	
Close to the critical point we can neglect the term $\langle \min \left(r_1,r_2\right)\rangle$, since it diverges slower than $\langle r\rangle^2$ and we are left with $\partial_f\langle r\rangle\simeq\langle r\rangle^2/2$. Solve this differential equation, we obtain	
\begin{equation}
\langle r\rangle =\frac{2}{f_c-f}+O\left(\frac{1}{\left(f_c-f\right)^2}\right).
\label{gamma}
\end{equation}

The divergence of average avalanche lifetime as $f\to f_c^-$ is given by $\langle s\rangle \sim \left(f_c-f\right)^{-\gamma}$ and the first moment of $P(s,f)$ obeys $\partial_f\langle s\rangle =\langle s^\mu\rangle\langle s\rangle$ \cite{PhysRevE.58.7141}. Therefore, in the critical region one has $\langle r\rangle =\langle s^\mu\rangle=\partial_f \left(\ln\langle s\rangle\right)=\gamma/\left(f_c-f\right)+O\left(\left(f_c-f\right)^{-2}\right)$. From Eq.~(\ref{gamma}) we conclude that in the saBS model $\gamma=2$. Using the scaling relation $\gamma=\frac{\left(2-\tau\right)}{\left(1+\mu-\tau\right)}$ \cite{PhysRevE.53.414}, we find in the saBS model $\gamma=2$ implies	
\begin{equation}
\tau=2\mu
\label{scaling1}.
\end{equation}

Asymptotically $r=As^\mu$, where $A$ is a constant, and $s_0^{1-\tau}\sim P\left(s>s_0\right)=P\left(r>As_0^\mu\right)\sim \left(As_0^\mu\right)^{1-\tau_r}$. So, in the $1d$ saBS model the exponent $\tau_r$ of the distribution of avalanche spatial sizes is related to the more familiar exponent $\tau$ of the distribution of their temporal durations through $\tau_r=\left(\tau-1\right)/\mu+1$. Combining this with the above exponent relation Eq.~(\ref{scaling1}), and using our best estimated value $\tau_r=1.299(2)$, we have $\tau=1.176(2)$, which is in good agreement with our simulation result $\tau=1.176(4)$.

\section{Conclusions}
\label{Conclusions}		
In conclusion, we have analyzed the critical behavior of the saBS model. This model is analyzed by Monte Carlo simulations and numerical integration of the master equation in one dimension. Nontrivial SOC state is observed for the saBS model with a nonzero interaction strength. In addition, we have shown that the saBS models with different interaction strengths exhibit the same critical behavior, i.e. they have same critical exponents and belong to the same universality class. The threshold fitness of the saBS model relies on the interaction strength. A stronger interaction strength will facilitate the propagation of mutation signal and thus leads to a lower threshold fitness. Finally, We have demonstrated that the nontrivial relation $\tau=2\mu$, derived for the aBS model, holds for its stochastic version as well.

\section*{Acknowledgments}
This work was in part supported by the Program of Introducing Talents of Discipline to Universities (Grant No. B08033) and National Natural Science Foundation of China (Grant No. 11505071).

\section*{Author contribution statement}
J.H. implemented the experiments and prepared all the figures. J.H., Z.S. and W.L. analyzed the results. All authors wrote, reviewed and approved the manuscript.

\bibliographystyle{epj}

\end{document}